\documentclass[prb,twocolumn,amsmath,amssymb,superscriptaddress]{revtex4}

\usepackage{graphicx}
\usepackage{dcolumn}
\usepackage{hhline}
\usepackage{bm,color}

\usepackage{float}

\usepackage{amsmath}

\usepackage{amsfonts}

\usepackage{amssymb}

\usepackage{epstopdf}

\begin{document}


\title{Relation between Kitaev magnetism and structure in $\alpha$-RuCl$_3$}

\author{A. Glamazda}
\affiliation{Department of Physics, Chung-Ang University, Seoul 156-756, Republic of Korea}

\author{P. Lemmens}
\affiliation{Institute for Condensed Matter Physics, TU Braunschweig, D-38106 Braunschweig, Germany}
\affiliation{Laboratory for Emerging Nanometrology (LENA), TU Braunschweig, D-38106 Braunschweig, Germany}

\author{S.-H. Do}
\affiliation{Department of Physics, Chung-Ang University,  Seoul 156-756, Republic of Korea}

\author{Y. S. Kwon}
\affiliation{Department of Emerging Materials Science, DGIST, Daegu 42988, Republic of Korea}

\author{K.-Y. Choi}
\email[]{kchoi@cau.ac.kr}
\affiliation{Department of Physics, Chung-Ang University,  Seoul 156-756, Republic of Korea}


\begin{abstract}
Raman scattering has been employed to investigate lattice and magnetic excitations of the honeycomb Kitaev material $\alpha$-RuCl$_3$ and its Heisenberg counterpart CrCl$_3$. Our phonon Raman spectra give evidence for a first-order structural transition from a monoclinic to a rhombohedral structure for both compounds. Significantly, only $\alpha$-RuCl$_3$ features a large thermal hysteresis, consistent with the formation of a wide phase of coexistence. In the related temperature interval of $70-170$~K, we observe a hysteretic behavior of magnetic excitations as well. The stronger magnetic response in the rhombohedral compared to the monoclinic phase evidences a coupling between the crystallographic structure and low-energy magnetic response. Our results demonstrate that the Kitaev magnetism concomitant with fractionalized excitations is susceptible to small variations of bonding geometry.
\end{abstract}

\maketitle

\section{Introduction}

The search for quantum spin liquids is one of the celebrated topics of current condensed matter physics. Quantum spin liquids are topologically ordered phases, in which  many-body spins are long-range entangled and evade all types of symmetry breaking even for temperatures down to $T=0$~K~\cite{Balents}. These phases feature exotic low-energy excitations including chargeless spinons emergent from spin fractionalization.  The exactly solvable Kitaev model on the honeycomb lattice is a particularly appealing system as it  harbors topological spin liquids and fractional quasiparticles, composed of itinerant Majorana fermion and localized flux~\cite{Kitaev,Baskaran,Knolle13,Knolle14}.

Jackeli and Khaliullin~\cite{JK} have provided the material requirements for realizing Kitaev interactions: strong spin-orbit coupled  Mott insulators with an edge-sharing octahedral environment where partially filled $t_{2g}$ orbitals are coupled via two 90$^{\circ}$ superexchange paths~\cite{JK,Chaloupka}.
To date, only a handful of candidates are known to fulfill  these criteria:
the harmonic-honeycomb iridates $\alpha$-, $\beta$-, and $\gamma$-Li$_2$IrO$_3$ ~\cite{Singh,Chun,Modic,Kimchi,Takayama,Glamazda16} and the $4d^5$ ruthenate $\alpha$-RuCl$_3$ ~\cite{Plumb,Majumder,Kubota,Weber,Sears,Sandilands,Nasu16,Banerjee,Banerjee2,Do,Kim,Rousochatzakis,Sandilands16,Yadav}.
All these materials, however, exhibit long-range magnetic ordering, preempting a spin liquid state
due to the presence of residual Heisenberg, off-diagonal, and longer-range interactions. Despite its magnetic ordering, the existence of Majorana quasiparticles was inferred from a broad continuum of excitations observed in inelastic light and neutron scattering measurements of $\alpha$-RuCl$_3$~\cite{Sandilands,Nasu16,Banerjee,Banerjee2,Do}. Furthermore, thermal conductivity and specific heat studies show a release of magnetic entropy and a coherent heat transport of itinerant quasiparticles at about 100~K, being consistent with the theoretical prediction for itinerant Majorana fermions~\cite{Hirobe}.

Notably, $\alpha$-RuCl$_3$ exhibits a strong variation in magnetic transition temperatures ranging from $T_\mathrm{N}=7-16$~K and in the number of the magnetic transitions, depending on synthesis conditions, sample quality and mechanical strain exerted on the samples. The observed multiple magnetic transitions are ascribed to different types of stacking faults between the weakly coupled honeycomb layers induced by van der Waals interaction~\cite{Banerjee}. The extreme sensitivity of $T_\mathrm{N}$ to the stacking of the layers suggests that the interlayer magnetic interactions rely on the stacking pattern. Nonetheless, the relation between the structural and magnetic subsystems is not particularly well understood. However, it could be a cornerstone in the understanding and interpretation of certain material properties.

At room temperature $\alpha$-RuCl$_3$ has a monoclinic $C2/m$ structure~\cite{Plumb,Majumder,Kubota,Sears,Johnson,Cao,Park}.
On the other hand, three different low-$T$ structures have been controversially reported: the monoclinic $C2/m$, the trigonal $P3{_1}12$, and the rhombohedral $R\bar{3}$ structure.  These structures differ in the stacking sequence of the honeycomb layers and in the relative displacement of neighboring layers and, importantly, in the bond distances and angles. Near the 90$^{\circ}$ bond geometry, the strength of Kitaev and Heisenberg interactions depend strongly on the Ru-Cl-Ru bond angles~\cite{Yadav}. The energy differences between these structures are very small and the experimental structural analysis is often hampered by the stacking faults~\cite{Kim,Johnson}.  It is remarkable that certain samples show a first-order structural phase transition to a low-temperature rhombohedral structure~\cite{Kubota,Park}. Presently experiments on single layer materials are still rather rare~\cite{Weber}.

As such, the comparison between $\alpha$-RuCl$_3$ and its sister compound CrCl$_3$ will shed light on investigating the relation between the structure and Kitaev magnetism because CrCl$_3$ having Heisenberg interactions undergoes the monoclinic-to-rhombohedral structure transformation at 240~K~\cite{Cable,Morosin}. In Kitaev materials, Raman spectroscopy presents a powerful experimental tool as it probes Majorana fermion density of states~\cite{Knolle14}. In addition, lattice and magnetic excitations can be simultaneously detected and their mutual couplings can be deduced from the spectral form of their Raman responses. In the earlier report, Sandilands {\it et al.}~\cite{Sandilands} found Raman spectroscopic signatures of the fractionalized excitation and Fano resonance in $\alpha$-RuCl$_3$. However, lattice anomalies related to the structural phase transition were not addressed, thereby disallowing to quantify their influence on magnetic excitations.

In this paper, we present a comparative Raman scattering study of $\alpha$-RuCl$_3$ and its isostructural counterpart CrCl$_3$. Both compounds are found to undergo first-order phase transitions from a monoclinic to a rhombohedral structure. Strikingly, we observe large thermal hysteresis in the structural transition temperatures only in a case of $\alpha$-RuCl$_3$. The wide temperature range with coexisting phases between $70-170$~K enables us to differentiate the magnetic response of the monoclinic and the rhombohedral structure. The stronger magnetic signal in the rhombohedral phase demonstrates that Kitaev magnetism accompanying fractionalized excitations is susceptible to a small variation of the crystallographic structure.

\section{Experimental Details}
Single crystals of $\alpha$-RuCl$_3$ and CrCl$_3$ were synthesized by a vacuum sublimation method. Commercial RuCl$_3$ and CrCl$_3$ powders (Alfa-Aesar) were thoroughly ground, and dehydrated in a quartz tube for two days. The sealed ampoules in vacuum were heated to 1080$^{\circ}$~C for $\alpha$-RuCl$_3$ and 850$^{\circ}$~C for CrCl$_3$. They are cooled down to 600$^{\circ}$~C  at a rate of 2$^{\circ}$~C/h after staying for 24 hours. The platelike crystals have a shinny surface with orientation parallel to the $ab$ plane. The phase purity and the stoichiometric composition of both $\alpha$-RuCl$_3$ and CrCl$_3$ were confirmed by EDX (Electron Dispersive X-ray).

The structural and magnetic properties of $\alpha$-RuCl$_3$ were characterized by thermodynamic, x-ray and neutron diffraction measurements~\cite{Do,Park,Leahy}.
Compared to the earlier studies, our crystal shows  a single transition at $T_\mathrm{N}=6.5$~K, the lowest temperature
reported so far. As the multiple transitions are linked to stacking faults of
the RuCl layers, a single transition implies a nearly uniform stacking pattern in a low-$T$ region. In addition, our x-ray measurements
unveil a first-order structural phase transition from the high-$T$ monoclinic to the low-$T$ rhombohedral structure below
60~K.

For Raman experiments, we used the crystals with dimensions of $3\times3\times1$ mm$^3$. The samples were installed into a He-closed cycle cryostat with a temperature range of $T=8-300$~K. The scattered spectra were recorded by using a triple spectrometer (Dilor-XY-500) and a micro-Raman spectrometer (Jobin Yvon LabRam) equipped with a liquid-nitrogen-cooled CCD (charge coupled detector). The Raman spectra were taken in a (quasi)backscattering geometry with the excitation line $\lambda=532$~nm of a Nd:YAG (neodymium-doped yttrium aluminum garnet) solid-state laser. Heating effects do not exceed 1 K. We use a Ne-lamp to calibrate the spectral position of the spectrometer.

\section{Results and discussion}

\subsection{Phonon Raman spectra of CrCl$_3$}
\begin{figure}
\label{figure1}
\centering
\includegraphics[width=8cm]{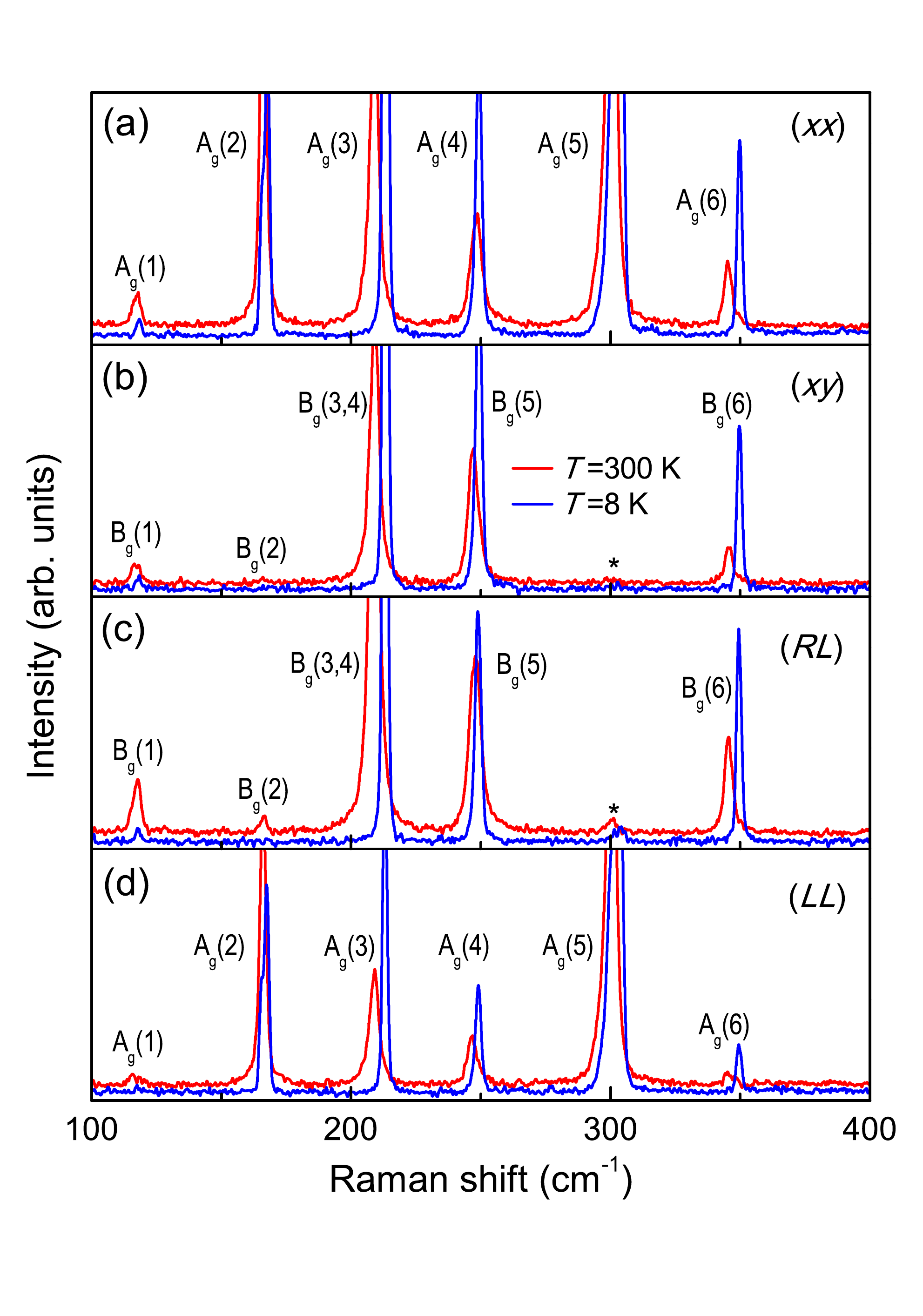}
\caption{(a),(b),(c) and (d) Comparison of Raman spectra of CrCl$_3$ at $T=8$ and 300~K in ($xx$), ($xy$), ($RL$) and ($LL$) polarizations, respectively. A$_g(i)$ and B$_g(i)$ ($i=1-6$) are the phonon modes allowed in each polarization. The asterisks indicate a weak forbidden A$_g$ mode showing up
in ($xy$) and ($RL$) polarizations due to a leakage of a polarizer.}
\end{figure}

In order to elucidate the anomalous structural and magnetic behaviors of $\alpha$-RuCl$_3$, we first focus on the sister compound CrCl$_3$, which undergoes a structural phase transition from the monoclinic structure ($C2/m$)
to the rhombohedral structure ($R\bar{3}$) at $T_\mathrm{S}=240$~K~\cite{Cable,Morosin}. Unlike $\alpha$-RuCl$_3$, CrCl$_3$ forms a conventional Heisenberg-type magnet having magnetic ordering at $T_\mathrm{N}=17$~K and thus its low-energy excitation is given by conventional spin wave without spin fractionalization~\cite{Cable}.

Figure 1 shows the polarized Raman spectra of CrCl$_3$ measured at $T=8$ and 300~K and in the in-plane ($xx$), ($xy$), ($RL$), and ($LL$) polarizations. The circular ($RL$) and ($LL$) polarizations correspond to $xx-yy-i(xy+yx)$ and $xx+yy-i(xy-yx)$ geometries with $R=x-iy$ and $L=x+iy$, respectively, allowing probing mainly the respective $\mbox{B}_g$ and $\mbox{A}_g$ symmetry channel at room temperature. In the high-$T$  monoclinic phase, we observe six  $\mbox{A}_g$ phonon modes at  117.4, 166.2, 209, 248.3, 300.4 and 345.3~cm$^{-1}$ in ($xx$) and ($LL$) polarizations and five $\mbox{B}_g$ modes at  117.4, 166.3, 209, 247.9 and 345.5~cm$^{-1}$ in ($xy$) and ($RL$) polarizations. The assignment of the observed phonon modes is based on the factor group analysis for the $C2/m$ space group, which predicts the following total irreducible representation for the Raman-active modes $\Gamma_{HT}=6\mbox{A}_g (aa, bb, cc, ac) + 6\mbox{B}_g (ba, bc)$. According to the lattice dynamic calculations~\cite{Kanesaka,Avram}, the two 209~cm$^{-1}$ modes  in the $\mbox{B}_g$ scattering channel are degenerate in energy, leading to one missing $\mbox{B}_g$ phonon. In addition, the $\mbox{A}_g$ and $\mbox{B}_g$ modes have almost same energies and thus the polarization dependence does not allow an unambiguous identification of the corresponding modes according to their energies. In the low-$T$ rhombohedral phase the Raman-active modes are factored as $\Gamma_{LT} =4\mbox{A}_g(aa, bb, cc)+ 4\mbox{E}_g (aa, bb, ab, ac, bc)$. Here the $\mbox{A}_g$ ($\mbox{B}_g$) mode of the high-$T$ phase is correlated with the $\mbox{A}_g$ ($\mbox{E}_g$) mode of the low-$T$ phase. The ($\mbox{A}_g$+$\mbox{B}_g$) mode of the high-$T$ phase is transformed into the double degenerate $\mbox{E}_g$ mode of the low-$T$ phase. As assigned in Fig.~2, at low temperatures we observe 4$\mbox{E}_g$ modes and 2$\mbox{A}_g$ modes (possibly due to a leakage of a polarizer)
in (RL) polarization~\cite{Bermudez}.

\begin{figure}
\label{figure2}
\centering
\includegraphics[width=9cm]{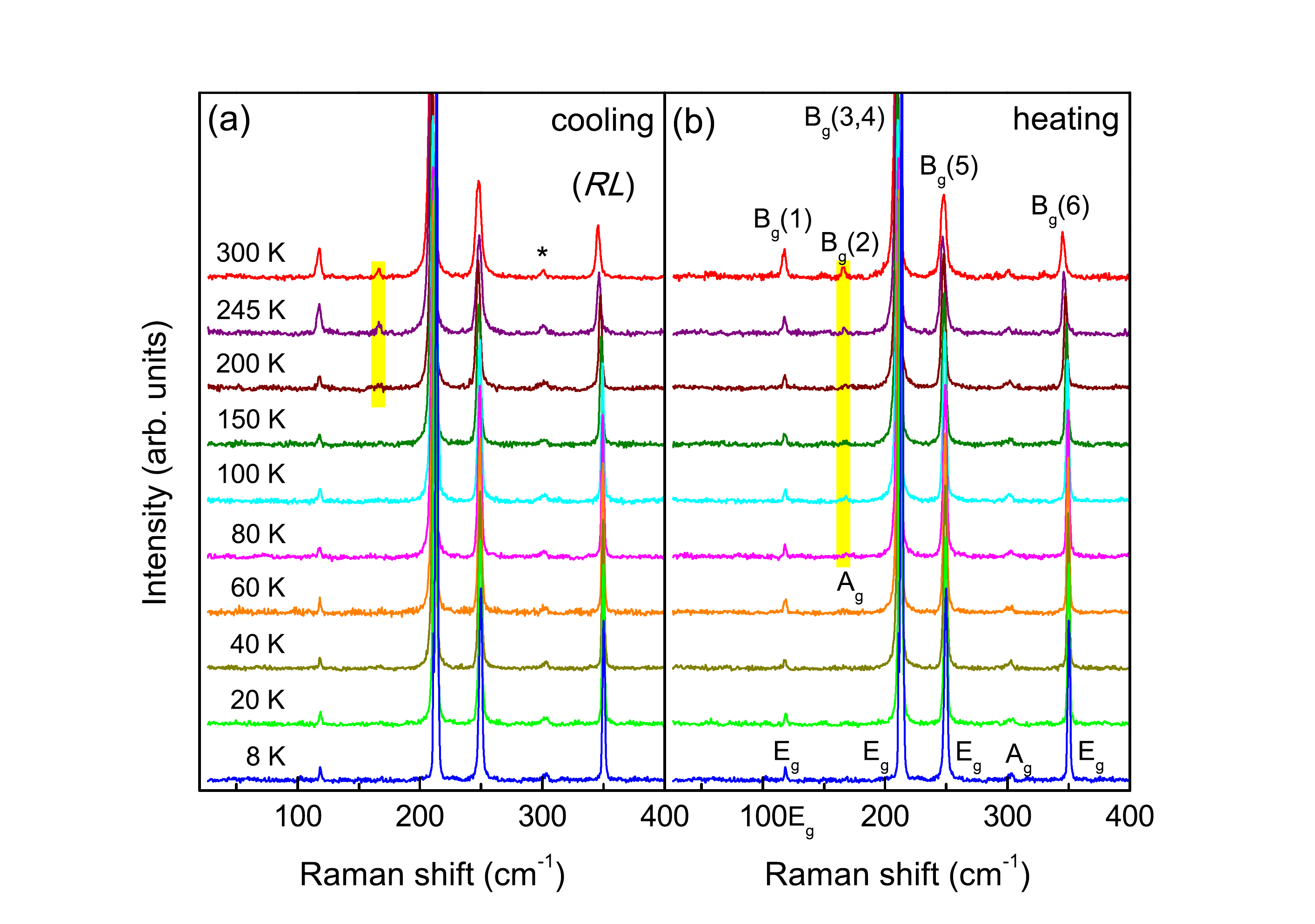}
\caption{Temperature dependent Raman spectra of CrCl$_3$ measured upon cooling (a) and heating (b) in ($RL$) polarization. The yellow bars mark the phonon which is present at a finite temperature interval.}
\end{figure}

To investigate the structural phase transition in CrCl$_3$ in detail, Raman spectra were measured between 300 and 8 K in ($RL$) polarization for each cooling and warming cycle. The results are compared in Fig.~2.  The advantage of employing the (RL) polarization is the suppression of low-frequency stray light. As the temperature is lowered below 240~K, the phonon modes exhibit no drastic changes in energy and intensity. This is due to the fact that the structural phase transition mainly involves shearing of the van der Waals bonded layers to change the relative translation between adjacent layers in the stacking sequence~\cite{McGuire}.
In the monoclinic structure, each adjacent layer is displaced along the $a$ direction while in the rhombohedral structure, the Cr in one layer is positioned directly on the center of the honeycomb lattice of the two neighboring layers. Given this consideration, no substantial differences are anticipated in intralayer bond character between the monoclinic and the rhombohedral phase. We note that the high-$T$ $\mbox{B}_g(2)$ mode disappears at 200~K (80~K) upon cooling (heating) as marked by the yellow vertical bars in Fig.~2.
This may be due to the polarizer leakage.

\begin{figure}
\label{figure3}
\centering
\includegraphics[width=8.5cm]{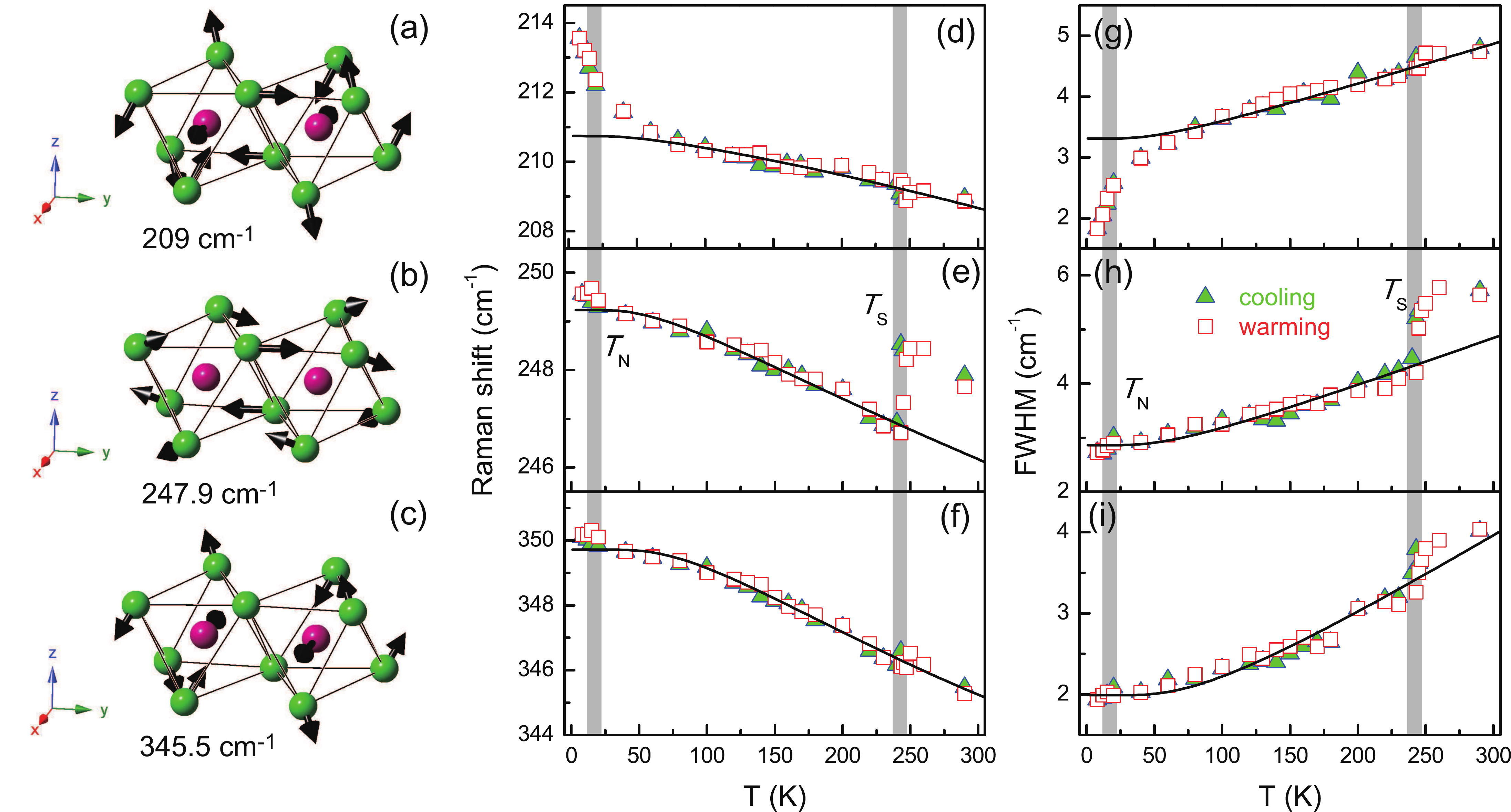}
\caption{A sketch of the eigenvectors of the (a) 209, (b) 247.9 and (c) 345.5~cm$^{-1}$ modes in CrCl$_3$. (d),(e),(f),(g),(h),(i) Temperature dependence of the frequency and full width at half maximum (FWHM) of the 209, 247.9 and 345.5~cm$^{-1}$ modes comparing cooling and warming. The vertical shaded bars indicate the magnetic and structural transition temperatures.  The solid lines are fits to the anharmonic model as described in the text.}
\end{figure}

For a quantitative analysis of phonon parameters in CrCl$_3$, the phonon spectra are fitted to a sum of Lorentzian profiles. The resulting frequencies and full width at half maximum (FWHM) for the 209, 247.9 and 345.5~cm$^{-1}$ B$_g$ modes are plotted in Fig.~3 as a function of temperature. The errors are smaller than the symbol size. A discontinuous change of the phonon parameters at $T_\mathrm{S}=240$~K indicates the first-order character of the structural transition. There exists only little thermal hysteresis between the warming and the cooling cycles.

Examining the phonon anomalies, the temperature dependence of the phonon frequency $\omega(T)$ and the FWHM $\Gamma(T)$
is described in terms of an anharmonic model~\cite{BW};
\begin{eqnarray}
\omega(T) =\omega_0 + A[1 + 2/(e^{\hbar\omega_0/2k_BT} - 1)],\\
\Gamma(T) =\Gamma_0+B[1 + 2/(e^{\hbar\omega_0/2k_BT} - 1)].
\end{eqnarray}
Here, $\omega_0$ and  $\Gamma_0$ are the bare frequency and the residual FWHM of the optical mode at $T=0$~K, respectively, and $A$ and $B$ are constants.
We find discernible deviations of the experimental data from the fitted curve on cooling through both $T_\mathrm{N}$ and $T_\mathrm{S}$ (see the solid lines in Fig.~3). As the magnetic order sets in, the phonon frequency shows an additional increase and the phonon linewidth an additional drop comparing with the anharmonic phonon-phonon estimation. The phonon hardening and narrowing occurring below $T_\mathrm{N}$ are ascribed to magnetoelastic coupling as reported in other Cr-based van der Waals materials~\cite{Casto,Tian}. The spin-phonon coupling induced anomalies are most pronounced for the 209~cm$^{-1}$ mode whose frequency and linewidth deviate from the conventional anharmonic model at $T\sim 70$~K.  This is associated with the onset of magnetic correlations developing at much higher temperatures than $T_\mathrm{N}$. The sensitivity of the 209~cm$^{-1}$ mode to spin-phonon coupling is rationalized by its normal mode displacement, which involves the modulation of the Cr-Cl-Cr bond strength and thus the superexchange interactions between the two Cr atoms as shown in Fig.~3(a).

In addition to the low-$T$ phonon anomalies in CrCl$_3$, the abrupt change of the phonon frequency and linewidth takes place upon cooling through $T_\mathrm{S}$.  The 247.9~cm$^{-1}$ mode shows an appreciable energy drop and line narrowing by about 2~cm$^{-1}$ while other modes exhibit marginal changes of the phonon energy  and linewidth. As sketched in Fig.~3(b), the 247.9~cm$^{-1}$ mode involves the out-of-phase vibration of Cl atoms mainly in the $ab$ plane and thus is closely tied to the shearing of the van der Waals bonded layers. This confirms that the crystallographic structure transition accompanies mainly a change in the stacking sequence. The 345.5~cm$^{-1}$ mode whose eigenvectors are sketched in Fig.~3(c) displays the weak anomalies at both $T_\mathrm{N}$ and $T_\mathrm{S}$.

\subsection{Phonon Raman spectra of $\alpha$-RuCl$_3$}

\begin{figure}
\label{figure4}
\centering
\includegraphics[width=7.5cm]{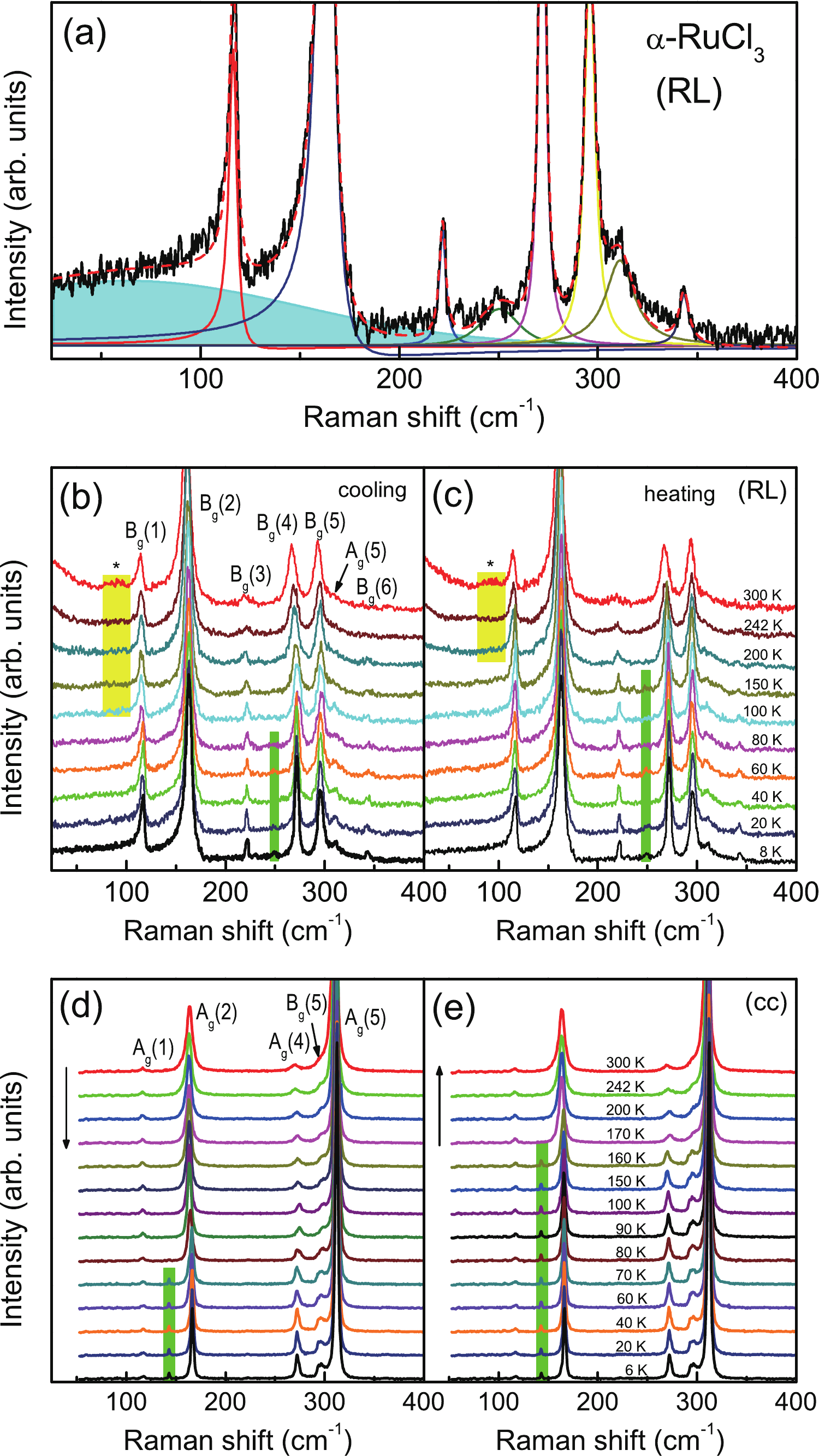}
\caption{(a) A fit of the $T=8$~K spectrum to a sum of a Gaussian profile (cyan shaded region), two Fano lines (red and blue solid lines), and six Lorentzian profiles (colored solid lines). The red dash line represents the total sum of a fitting.
Temperature dependence of the Raman spectra of $\alpha$-RuCl$_3$ measured on cooling (b) and heating (c) in the in-plane ($RL$) polarization as well as on cooling (d) and heating (e) in the out-of-plane (cc) polarization. The yellow bars mark the phonon which disappears below $T_\mathrm{S}$ and the green bars indicate the activated phonons below $T_\mathrm{S}$. }
\end{figure}

Having established the monoclinic-to-rhombohedral structure transition in CrCl$_3$, it has to be answered whether $\alpha$-RuCl$_3$ undergoes the same type of structural transformation. Indeed, a x-ray diffraction study of our crystal exhibits the occurrence of a first-order structure transition
from the monoclinic to the rhombohedral structure with large thermal hysteresis~\cite{Park}.
In the following we will employ optical phonons to investigate a structural transformation and
its coupling to magnetic excitations. For this purpose, we performed the Raman scattering measurements in the in-plane (RL) and the out-of-plane (cc) polarizations on cooling and on a subsequent warming run. The (RL) and (cc) polarizations probe the $B_g$ and $A_g$ symmetry at high temperatures, respectively. The corresponding Raman spectra as  a function of temperature are plotted in Fig.~4.

In (RL) polarization, we observe a total of eight modes both at room temperature and at $T=8$~K as shown in Figs.~4(a)-(c). Out of them, the six modes at 115.9, 161.3, 221.4, 268,5, 294.8 and 338~cm$^{-1}$ are assigned to $B_g$ modes by referring to the polarized Raman spectra of CrCl$_3$. Noteworthy is that the corresponding phonon frequencies between CrCl$_3$ and $\alpha$-RuCl$_3$ differ by only $1- 10$~cm$^{-1}$, indicating that the difference in their bonding strengths is not substantial. The 312.4~cm$^{-1}$ mode is assigned to the $A_g(5)$ modes since the intense phonon peak with the same energy shows up in the (cc) polarization spectra. This mode appears due to a slight tilting of the aligned crystal toward the $c$ axis. Noticeably, the low-frequency 88~cm$^{-1}$ mode which disappears below 100~K is not part of the Raman-active modes predicted for the C2/m space group. For a trigonal $P3{_1}12$ structure, the factor group analysis yields a total of 34 Raman-active modes; $\Gamma =11\mbox{A}_1(aa, bb, cc)+ 23\mbox{E}(aa, bb, ab, ac, bc)$. The large difference of the predicted and observed phonon modes makes a symmetry reduction to the trigonal structure improbable. Rather, the extra 88~cm$^{-1}$ mode may be ascribed to an infrared-active mode activated by local distortions and strains.

For temperatures below $T_\mathrm{S}$, we observe also a symmetry-forbidden phonon at 144~cm$^{-1}$ in (cc) polarization and at 249~cm$^{-1}$ in (RL) polarization. These activated phonons arise from reverse-obverse twinning and chemical disorder in the rhombohedral phase as a consequence of
stacking faults. Usually, local lattice distortions can activate symmetry-forbidden signals without inducing a global structural change.
Thus, the appearance of the extra phonon modes at both the low- and high-$T$ phase does not necessarily mean the symmetry reduction from
the low-$T$ rhombohedral and high-$T$ monoclinic structure, respectively. It is noted that such activated modes are lacking in CrCl$_3$.

As the temperature is lowered, in the intralayer polarization the low-frequency quasielastic response evolves into the broad continuum and the 116.6 and 163.7~cm$^{-1}$ modes become asymmetric in their lineshape, typical for Fano-type coupling
between the discrete optical phonons and the continuum excitation.
The respective phonons become symmetric and the broad continuum is no longer detectable in the interlayer polarization. The observation of the broad continuum and Fanon resonance only in the in-plane polarization is not compatible with a disorder-induced phonon background.
Rather, its specific polarization and temperature dependences demonstrate that the broad continuum is of magnetic origin
and is associated with two-dimensional Kitaev magnetism (see the section III. C).

\begin{figure}
\label{figure5}
\centering
\includegraphics[width=8cm]{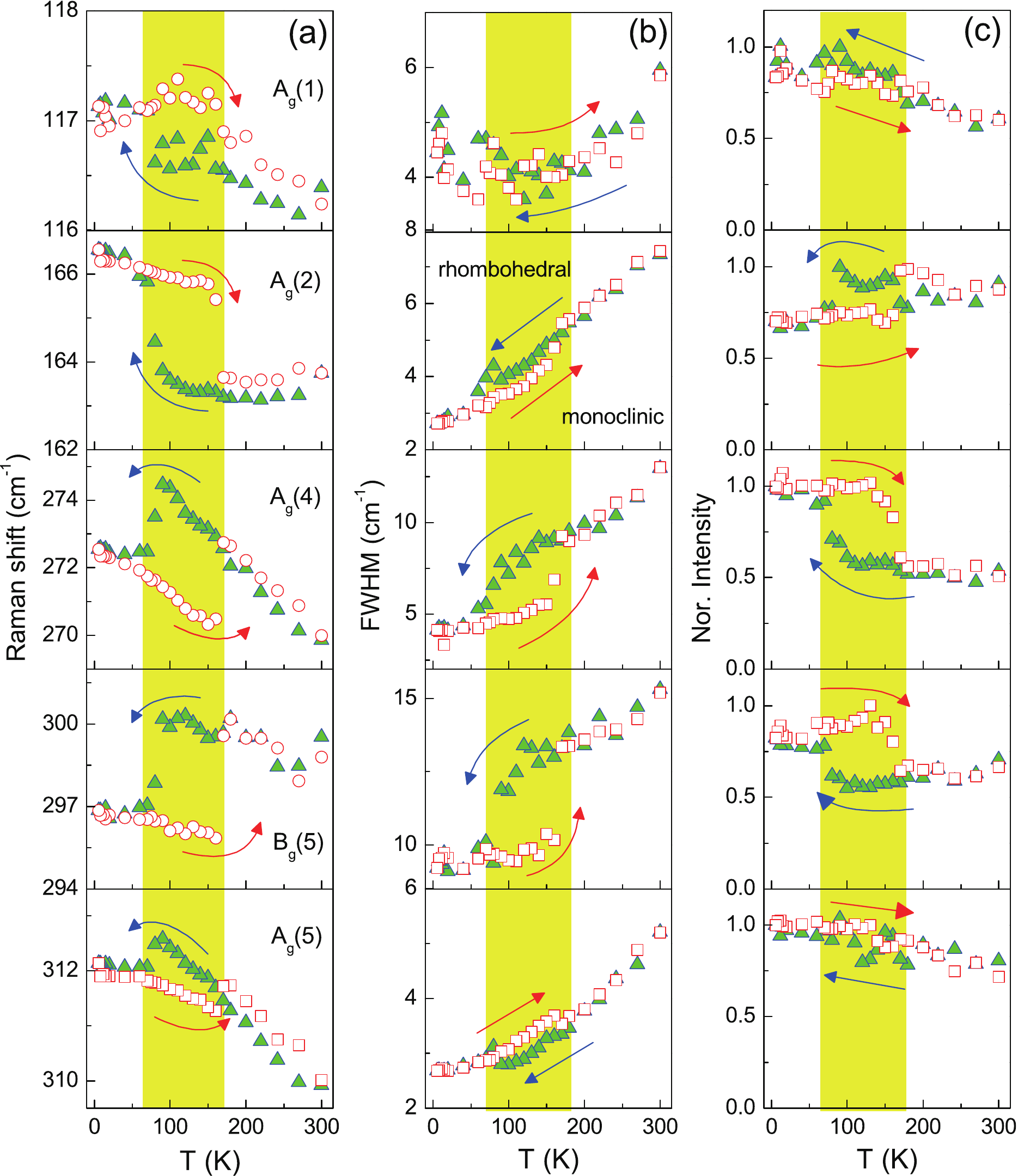}
\caption{Thermal hysteresis behavior of (a) the frequency, (b) the FWHM, and (c) the normalized intensity of the 116.4, 163.8, 270, 299.5 and 310~cm$^{-1}$ modes. The shaded region indicates the coexistence of the monoclinic and rhombohedral phase. The arrows indicate the direction of the temperature sweep.}
\end{figure}

Shown in Fig.~5 is the frequency, FWHM, and normalized intensity of the 116.4, 163.8, 270, 299.5 and 310~cm$^{-1}$ modes measured in each cooling and warming procedure. A drastic change of the phonon parameters is observed at $T_\mathrm{S}=70$~K on cooling. The transition increases to a significantly higher temperature  $T_\mathrm{S}=170$~K on the next warming cycle. The thermal hysteresis with width $\Delta T=100$~K corroborates that the structural transition is first order~\cite{Glamazda}.
On cooling through $T_\mathrm{S}$, the low-frequency 116.4 and 163.8~cm$^{-1}$ modes display a hardening by $1-3$~cm$^{-1}$, a narrowing by 1~cm$^{-1}$, and a sudden drop of the normalized intensity. This is contrasted by the intermediate-frequency 270 and 299.5~cm$^{-1}$ modes showing a softening by $3-4$~cm$^{-1}$, a narrowing by 3~cm$^{-1}$, and a strong increase of the normalized intensity. The 310~cm$^{-1}$ mode is weakly affected by the structural change. The disparate impacts of the structural transformation on the phonon energies, lifetimes, and intensities imply that the interlayer and intralayer bond characters change in a different manner through $T_\mathrm{S}$, judging from the fact that the low-frequency modes
contain out-of-phase motions of Ru-Cl-Ru bonds while the intermediate-frequency modes out-of-phase motions of Cl atoms along the out-of-plane direction (see below for further discussions).

We recall that for the case of CrCl$_3$, the 248.3~cm$^{-1}$ mode exhibits a similar trend of the phonon anomalies through $T_\mathrm{S}$ but other modes are largely intact. In addition, the thermal hysteretic behavior is hardly visible. This is in stark contrast to $\alpha$-RuCl$_3$ showing a wide range of temperatures over which the high- and low-temperature phases coexist. The large thermal hysteresis and
the occurrence of the activated phonons in $\alpha$-RuCl$_3$ indicate that $\alpha$-RuCl$_3$
is more vulnerable to stacking faults than CrCl$_3$ possibly due to weaker interlayer interactions.

\subsection{Spin fractionalization and spin-phonon coupling}
\begin{figure}
\label{figure6}
\centering
\includegraphics[width=8cm]{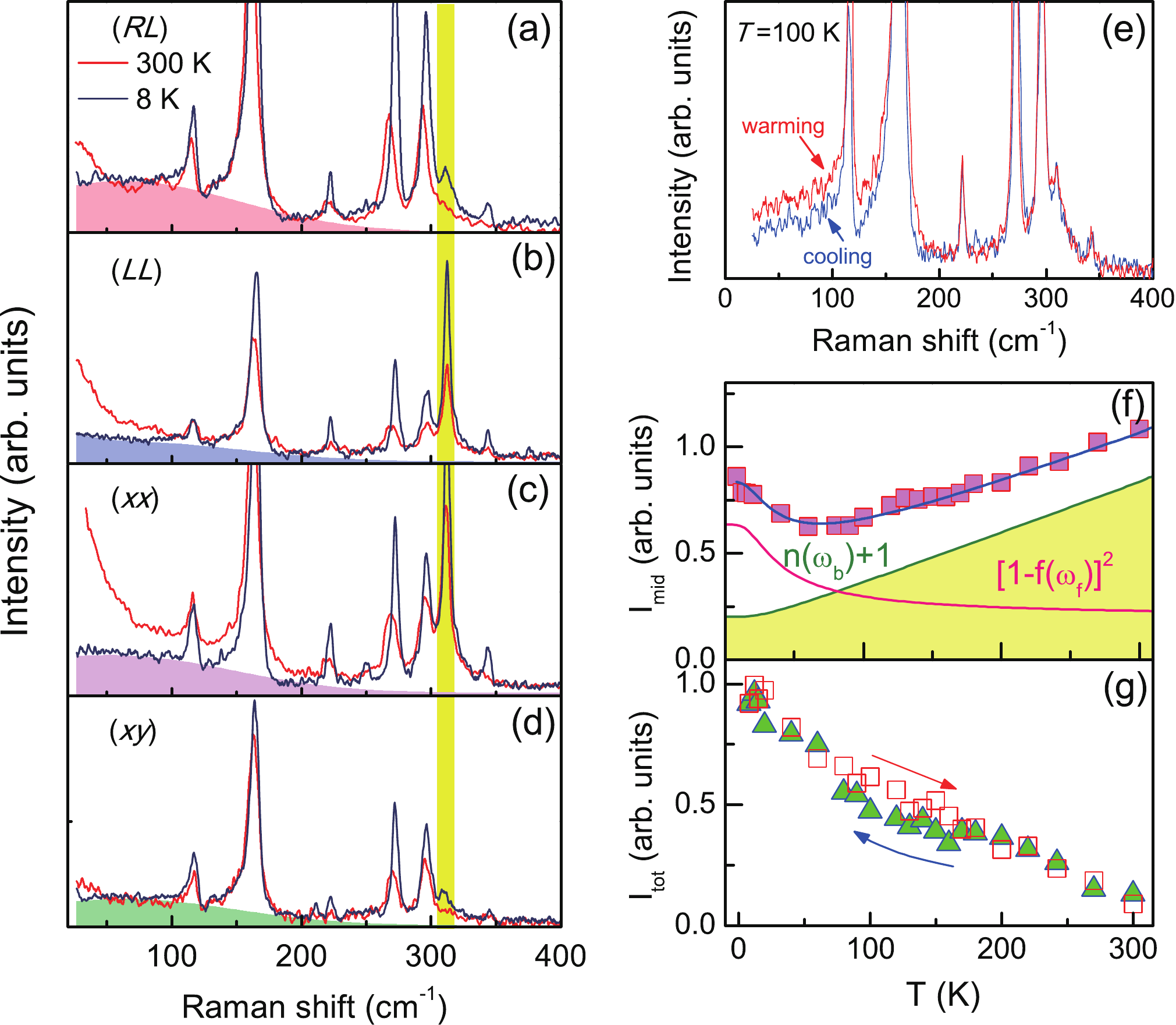}
\caption{(a),(b),(c),(d) Comparison of the magnetic Raman spectra among four different (RL), (LL), (xx), and (xy) scattering polarizations measured at $T=8$ and 300~K.  A magnetic continuum of each data set is emphasized by a color shading. The yellow bar marks the phonon which shows the strongest variation of its intensity with polarization. (e) Comparison
of the $T=100$~K Raman spectrum between the warming and cooling runs. (f) Temperature dependence of the integrated Raman intensity $I_{\mathrm{mid}}$ obtained in the middle energy window from 40 to 120~cm$^{-1}$. The shaded area indicates the bosonic background and the solid red line is a fit to a temperature dependence expected for a two-fermion creation or annihilation process, $[(1-f(\omega_f)]^2$ with the Fermi distribution function $f(\omega_f)$. (g) Thermal hysteresis of the total intensity  obtained by the integration of a magnetic continuum
up to 250 ~cm$^{-1}$.}
\end{figure}

Examining the effect of the structural transformation on a magnetic response, we focus on the magnetic continuum of $\alpha$-RuCl$_3$ measured at $T=8$ and 300~K in the (RL), (LL), (xx), and (xy) scattering symmetries. The obtained Raman spectra are compared with respect to polarization in Figs.~6(a)-(d). We observe a broad magnetic continuum with a residual spectral weight at low frequencies (marked by color shadings), which is in accordance with the previous work of Ref.~\cite{Sandilands} with respect to spectral shape and energy range. In a pure Kitaev system, the continuum excitation originates from two-Majorana scattering~\cite{Knolle14,Nasu16}. In the presence of perturbation terms, however, it is far from clear to what extent the coherent Majorana fermion contributes to the magnetic continuum. The ratios of spectral weight comparing different polarizations are given by I(xx):I(xy):I(RL):I(LL)=0.78:0.57:1.0:0.46. A detailed $T$-dependence study was made in (RL) polarization corresponding to the $E_g$ channel by exploiting the strongest intensity of its magnetic continuum. As compared in Fig.~6(e), we can identify a noticeable difference of the
magnetic continuum between the cooling and warming runs below 200~cm$^{-1}$ at $T=100$~K, in the coexisting phase.

To identify the Majorana fermion contribution to the magnetic continuum, the magnetic Raman intensity in (RL) polarization is integrated over the middle energy range
of $40 <\omega<120$~cm$^{-1}$ and its temperature dependence $I_{\mathrm{mid}}(T)$ is plotted in Fig.~6(f). In the chosen energy interval, a calculated magnetic Raman intensity for the ideal Kitaev honeycomb lattice shows that
the scattering process is dominated by the creation or annihilation of pairs of Majorana fermions~\cite{Nasu16}. As a result, $I_{\mathrm{mid}}(T)$ is described asymptotically by the two-fermion scattering form $[(1-f(\omega_f)]^2$ with $f(\omega_f)=1/(1+ e^{\hbar\omega_f/k_BT})$ and fermion energy $\omega_f$. In a real compound, however, the bosonic scattering should be invoked because
the residual interactions give rise to incoherent magnetic excitations including correlated magnons.  Thus, $I_{\mathrm{mid}}(T)$ is given by a sum of the Bose factor $1 + n(\omega_b) = 1/[1 - \exp(-\hbar\omega_b/k_BT )]$ with $\omega_b=49$~cm$^{-1}$ and the two-fermion scattering contribution $[(1-f(\omega_f)]^2$ with $\omega_f=42$~cm$^{-1}$. In this asymptotic formulation, $\omega_b$ and $\omega_f$ are associated with the effective energies of the fermionic and bosonic excitations. Essentially the same quasiparticle excitations with  $\omega_f=0.62~J (=80~\mathrm{cm}^{-1})$ were extracted from the numerical Raman intensity integrated over the similar energy range $0.5 J <\omega<1.5 J$ by Nasu et al.~\cite{Nasu16}. We note that the  asymptotic $[(1-f(\omega_f)]^2$ behavior remains valid under a small variation of
the middle frequency window. Taken together, the magnetic continuum comprises the coherent Majarana fermions and the incoherent magnetic excitations. The observation of the Y-shaped high-energy dispersive excitation by a recent inelastic neutron scattering
study~\cite{Do} supports this interpretation.

We further calculate the total intensity $I_{\mathrm{tot}}(T)$ by the integration of the whole magnetic continuum up to $3.5J$.
The resulting intensity is plotted in Fig.~6(g) on both cooling and warming cycles. In contrast to $I_{\mathrm{mid}}(T)$,
no asymptotic form is known for $I_{\mathrm{tot}}(T)$, which contains a combination of creation and annihilation of
multiple Majorana fermions. $I_{\mathrm{tot}}(T)$ increases with lowering temperature in a monotonic manner. A close inspection reveals a small but discernible hysteresis: in the temperature interval of coexistence, $I_{\mathrm{tot}}(T)$ is slightly stronger on warming than cooling. This is a clear manifestation of coupling of the crystal structure to the Kitaev magnetism, implying that the itinerant Majorana fermions are more well-defined in the rhombohedral (open squares) than the monoclinic structure (full triangles).  A coupling between magnetism and lattice structure is a characteristic feature of strong spin-orbit coupled materials~\cite{Winter}.
In the rhombohedral structure, the three bonds are identical with a Ru-Cl-Ru angle of nearly $94^{\circ}$, forming isotropic Kitaev interactions~\cite{Park}. Remarkably, quantum chemistry calculations~\cite{Yadav} show that the nearest-neighbor Kitaev interaction $K$ is strongest around this angle while the nearest-neighbor Heisenberg interaction $J$ is weakest. Thus, the rhombohedral phase of $\alpha$-RuCl$_3$ with a large ratio of $\mid K/J \mid$ is close to the 2D Kitaev honeycomb model.
In the monoclinic phase, the bond angles deviate from  $94^{\circ}$ by a few degrees~\cite{Park}, rendering the Kitaev interactions anisotropic and the $\mid K/J \mid$ ratio smaller. Consequently, a small deviation from the $94^{\circ}$ bond angle tunes the system further away from the spin liquid phase.

\begin{figure}
\label{figure7}
\centering
\includegraphics[width=8cm]{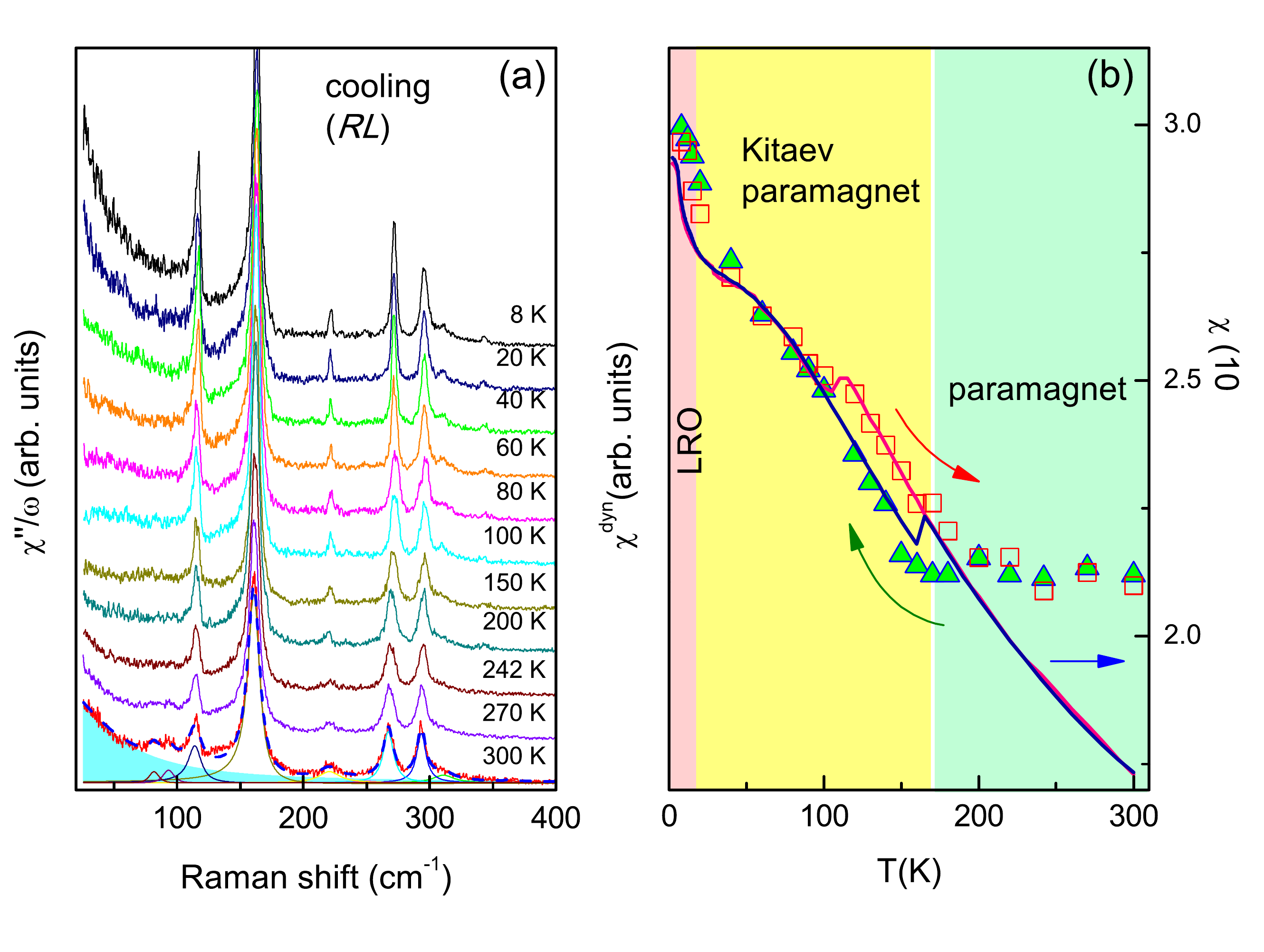}
\caption{(a) Temperature dependence of the Raman conductivity $\chi''/\omega$ on cooling in (RL) polarization. The cyan shading is a magnetic continuum.
(b) Temperature dependence of the dynamic Raman susceptibility deduced from $\chi''/\omega$ using the Kramers Kronig relation. Temperature dependence of the static spin susceptibility for $\mu_0H//c$ is plotted together for comparison.}
\end{figure}

In the further exploration of the Majorana fermions, we evaluate the dynamic spin susceptibility from the magnetic continuum. For this analysis,
we first obtain the Raman response  $\chi''(\omega)$ from the raw Raman spectra $I(\omega)$ using the relation  $I(\omega) \propto  [1 + n(\omega)]\chi''(\omega)$ and, therefrom, we define the Raman conductivity $\chi''(\omega)/\omega$. The temperature dependence of $\chi''/\omega$ on cooling is shown in Fig.~7(a). The Raman conductivity features a pronounced peak at $\omega=0$. Its amplitude varies strongly with temperature. Using the Kramers-Kronig relation $\chi^{\mathbf{dyn}}\equiv\frac{2}{\pi}\int^{\infty}_{0}\frac{\chi''(\omega)}{\omega}d\omega$, the dynamic spin susceptibility $\chi^{\mathbf{dyn}}$ is calculated~\cite{Gupta,Gallais}. In doing that, $\chi''(\omega)/\omega$ was first extrapolated for frequencies down to 0~cm$^{-1}$ and then integrated over the frequency range of $0-350$~cm$^{-1}$. Figure~7(b) plots the temperature dependence of $\chi^{\mathbf{dyn}}(T)$ together with the static magnetic susceptibility $\chi^{\mathbf{stat}}(T)$ for $\mu_0H//c$. Both $\chi^{\mathbf{dyn}}(T)$ and $\chi^{\mathbf{stat}}(T)$ demonstrate a rapid decrease with appreciable temperature hysteresis in the phase of coexistence, confirming that the magnetism and crystallographic structure are tied to each other. In the respective temperature interval, $\chi^{\mathbf{dyn}}$ is larger on warming (open squares; rhombohedral phase) than on cooling (full triangles; monoclinic phase). Again, this supports the notion that the itinerant Majorana fermions are somewhat more pronounced in the rhombohedral than in the monoclinic phase.

A reasonable agreement between $\chi^{\mathbf{dyn}}(T)$  and $\chi^{\mathbf{stat}}(T)$ is found in the temperature interval of $10- 170$~K. Clear deviations are visible in the low-$T$ zigzag ordered state and the high-$T$ paramagnetic state where spins are rigid and uncorrelated, respectively. The characteristic temperature $T=170$~K may correspond to a crossover from a simple paramagnet to a Kitaev paramagnet, at which spins start to be fractionalized to itinerant Majorana fermions due to the development of short-range
spin correlations~\cite{Nasu}. Indeed, the recent inelastic neutron scattering measurements evidence the emergent Majorana fermions below $120-130$~K~\cite{Banerjee2,Do}.  We further note that this crossover temperature coincides with the structural phase transition temperature on warming. It is likely to invoke magnetoelastic coupling as its origin. This is more than a coincidence since in the family of Cr trihalies crystallographic phase transitions are observed in a similar range of temperatures~\cite{McGuire}. Therefore we conclude that the interlayer interactions provide a driving force for stabilizing the high-symmetry rhombohedral phase at low temperatures.

\begin{figure}
\label{figure8}
\centering
\includegraphics[width=8cm]{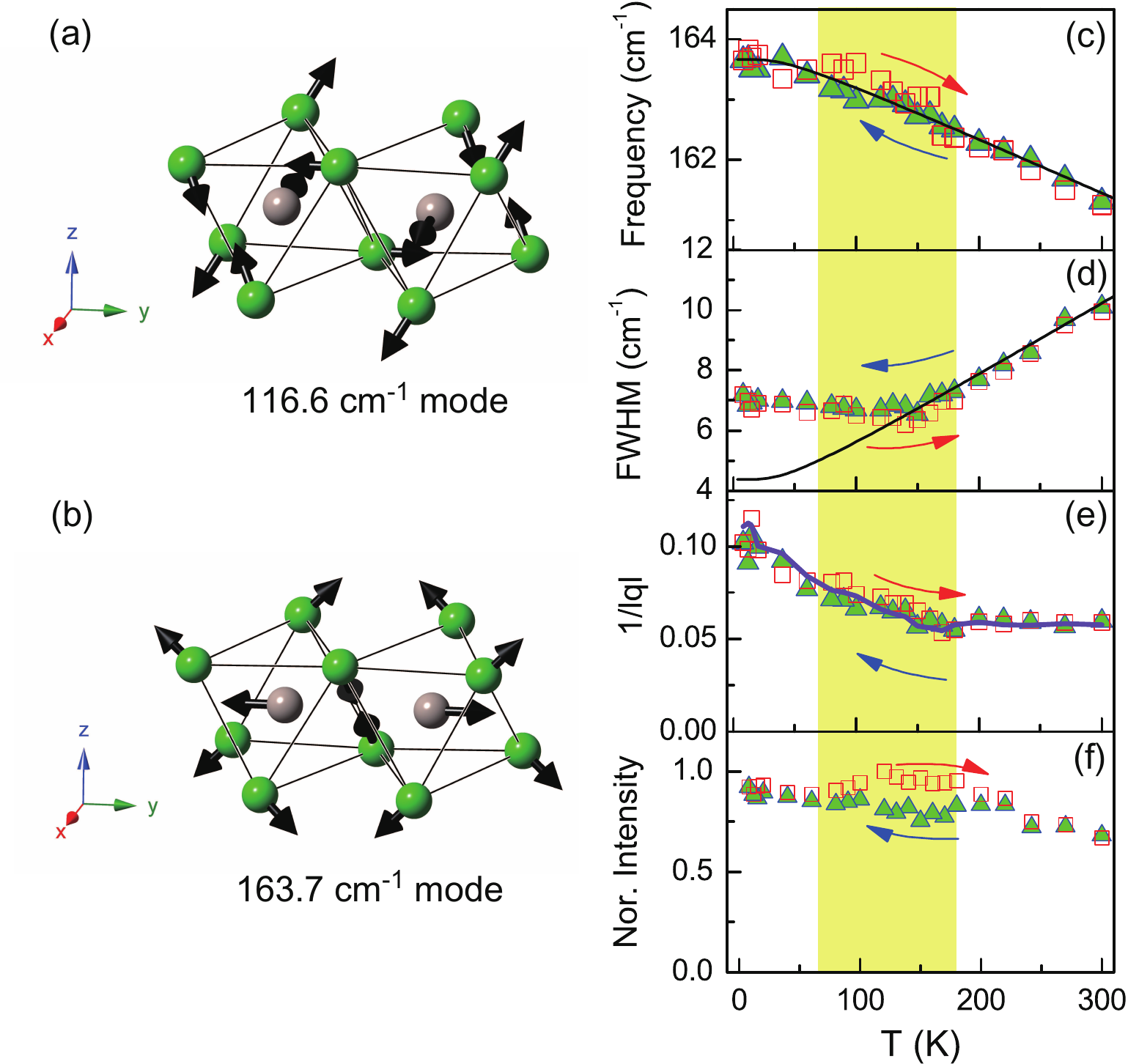}
\caption{A schematic representation of eigenvector of (a) the 116.6 and (b) the 163.7 cm$^{-1}$ modes of the monoclinic phase. The amplitude of the vibrations is represented by the arrow length. Gray balls indicate Ru ions and green balls are Cl ions. Temperature dependence of (c) the frequency, (d) the linewidth, (e) the Fano asymmetry $1/\mid q\mid$, and (f) the normalized intensity. The thin solid lines are a fit to an anharmonic phonon model and the thick solid line is the dynamic spin susceptibility
taken from Fig.~7(b).}
\end{figure}

Optical phonons can provide further information on the magnetic continuum as it is coupled to discrete lattice modes. Indeed, the asymmetric lineshape of the 116.6 and 163.7~cm$^{-1}$ optical phonons is clear evidence for a Fano resonance [see Fig.~4(a)]. As sketched in Figs.~8(a) and 8(b), these low-frequency modes entail out-of-phase motions of Ru atoms along the $a$ and $b$ axis together with stretching and bending vibrations of RuCl$_6$ octahedra, respectively. Thus, these lattice vibrations modulate directly the bonds mediating the Kitaev interaction.

For a quantitative analysis, both phonons are fitted to a Fano profile, $I(\omega) = I_0(q +\epsilon)^2/(1+\epsilon^2)$, where a reduced energy defined by $\epsilon = (\omega -\omega_0)/\Gamma$, $\omega_0$ is the bare phonon frequency, $\Gamma$ is the linewidth, and $q$ is the asymmetry parameter.  In Figs.~8(c)-(g), the resulting frequency shift, the linewidth, the Fano asymmetry, and the normalized intensity are summarized for the 163.7~cm$^{-1}$ mode. The errors are within a symbol size. The 116.6~cm$^{-1}$ mode exhibits the same trend and is thus omitted here.

The temperature dependence of the phonon frequency is captured by the anharmonic lattice model given by Eq.~(1) except for the small anomaly on warming in the temperature interval of coexisting phases [see Fig.~8(c)]. As evident from Fig.~8(d), the anharmonicity fails to describe the linewidth for temperatures below 140~K. The additional line broadening points towards the existence of another relaxation channel in addition to anharmonic phonon processes. Figure~8(f) shows that the Fano asymmetry parameter $1/\mid q\mid$ starts to increase progressively on cooling down below 160~K. A salient feature is that the Fano asymmetry parameter goes hand in hand with the dynamic magnetic susceptibility [compare the solid line and the symbols in Fig.~8(f)]. This implies that the Fano lineshape is an indicator of spin fractionalization to the Majorana quasiparticles. The same observation is reported for the harmonic-honeycomb iridates $\beta$-, and $\gamma$-Li$_2$IrO$_3$~\cite{Glamazda16}. The increasing $1/\mid q\mid$ with decreasing temperature can be translated to a growth of spin fractionalization with the onset temperature of 170~K. As shown in Figs~8(c) and (f), the phonon frequency and its normalized intensity exhibit a detectable hysteric behavior in the coexisting phase. In contrast, the phonon linewidth and $1/\mid q\mid$ do not display a noticeable thermal hysteresis. This seemingly contradictory behavior  may be related
to an intriguing interplay of two factors in determining the phonon parameters of the 163.7~cm$^{-1}$ mode:
(i) the structural modification through $T_\mathrm{S}$ and (ii) the slightly different coupling of the phonon
to the magnetic continuum between cooling and warming cycles.

\section{Conclusion}
To conclude, we have presented a comparative study of the isostructural Kitaev $\alpha$-RuCl$_3$ and Heisenberg material CrCl$_3$ with view of elucidating
the distinct crystallographic and magnetic properties of $\alpha$-RuCl$_3$. The compounds both undergo a first-order structural transition from a high-temperature monoclinic to a low-temperature rhombohedral phase. The difference is observed in the presence of a large hysteresis of the structural transitions on cooling and warming only for $\alpha$-RuCl$_3$. This structural peculiarity allows differentiating a magnetic response between the monoclinic and the rhombohedral phase. In the intermediate temperature range with coexisting phases the magnetic excitations are somewhat more pronounced on warming compared to cooling. Our experimental findings demonstrate that the rhombohedral structure has a more optimal bond geometry to host Kitaev magnetism than the monoclinic structure. Thus, it will be of great interest to investigate further whether a Kitaev liquid phase can be achieved through engineering the bond geometry by applying chemical and hydrostatic pressure in ruthenium trihalides.

\section*{Acknowledgment}
We thank  S.-Y. Park,  S. Ji and J.-H. Park for fruitful discussions. This work was supported by the Korea Research Foundation (KRF)
grant funded by the Korea government (MEST) (Grant No. 2009-0093817 and No. 2016-911392) as well as by German-Israeli Foundation (GIF, 1171-486 189.14/2011), the NTH-School Contacts in Nanosystems: Interactions, Control and Quantum Dynamics, the Braunschweig International Graduate School of Metrology, and DFG-RTG 1953/1, Metrology for Complex Nanosystems.
Y.S.K acknowledges the support from the Ministry of Science, ICT and Future Planning (Grant No. 2015M2B2A9028507).

\section*{References}


\end{document}